\documentstyle[preprint,aps]{revtex}

\begin{document}

\preprint{\vbox{\baselineskip=15pt
\hbox{NSF-ITP-93-90} \hbox{gr-qc/9308001} \hbox{August 2, 1993}}}

\title{Bianchi Cosmologies: New Variables \\
and a Hidden Supersymmetry}

\author{Octavio Obreg\'on\cite{th1}}
\address{Apartado Postal E-143, C. P. 37150 \\ Le\'on, Guanajuato,
M\'exico}

\author{Jorge Pullin }
\address{Department of Physics, University of Utah,
Salt Lake City, UT 84112\\
and\\
Center for
Gravitational Physics and Geometry,\\
The Pennsylvania State University,
University Park, PA 16802. }

\author{Michael P. Ryan}
\address{ Instituto de Ciencias Nucleares, UNAM\\
Apartado Postal 70-543 M\'exico 04510 D. F., M\'exico.}

\maketitle
\begin{abstract}
We find a supersymmetrization of the Bianchi IX cosmology in terms of
Ashtekar's new variables. This provides a framework for connecting the
recent results of Graham and those of Ryan and Moncrief for quantum
states of this model. These states are also related with the states
obtained particularizing supergravity for a
minisuperspace. Implications for the general theory are also briefly
discussed.
\end{abstract}

\section{Introduction}

Due to our inability to manage the canonical quantization of
the full equations of General Relativity in the generic case,
the minisuperspace approximation has been used several times to
find results in the hope that they would illustrate behaviors of
the general theory. The Bianchi cosmologies are the prime example.
Even in this simplified case little progress has been achieved for
the more generic model, the Bianchi IX cosmology. In fact, the
classical model shows various signs of chaotic behavior \cite{PuBeDeRo} and
this has usually been assumed to imply important complications for
the quantization of the model. In particular, until just recently,
not a single solution of the Wheeler-DeWitt (WDW) equation for the
Bianchi IX cosmology was known, resembling the situation that
one faces in the full theory.

In the years, the introduction of the Ashtekar
variables \cite{As} led to a change in this situation.
Recently, Kodama \cite{Ko} formulated Bianchi cosmologies in terms
of these variables and found a solution to the WDW
equation for the Bianchi IX case.  The meaning of this possible
quantum state remains unclear. Another interesting observation was
made by Moncrief and Ryan \cite{MoRy}. They pointed out that in terms of
the Ashtekar formulation (in a certain factor ordering), $\Psi = 1 $
was a possible solution of the WDW equation. The point is that when
translated into the traditional variables, this state had the form
$\exp (-I)$, where $I$ is a solution to  the corresponding Euclidean
Hamilton-Jacobi equation (with the sign of the potential changed).
They explicity checked that it solved
the WDW equation, providing the first known example of a solution
of the WDW equation in the traditional variables.

Almost simultaneously, Graham \cite{Gr} found a closely related state using
a very different technique. Exploiting what he calls a ``hidden
symmetry'' (a point that will be discussed below) in the WDW equation
he was able to supersymmetrize the model. The supersymmetry equations
are complicated to solve for the full wavefunction. However,
restricted to the bosonic sector, they yield a solution to the WDW
equation.

On the other hand, in supergravity the Bianchi class A
models seem to have the same kind of state as was pointed out by
D'Eath, Hawking and Obreg\'on \cite{DeHaOb} and for the Bianchi IX
case D'Eath
\cite{De} proved that only the term $\exp (-I)$ is permitted in the wave
funtion. This behaviour is, as expected, present in the other Bianchi
class A models, \cite{AsTaYo}.

The three previous procedures virtually result in the same quantum
state. This state appears as related to the wormhole ground state
\cite{DeHaOb,De,HaPaVi}.
It is also present in the full theory of supergravity $N=1 $
\cite{De2}.  The discussion of these points however, lies beyond the
scope of this paper.

The main point of this paper is to point out a connection between the
results of Moncrief and Ryan and those of Graham. We will show how one
can supersymmetrize the Bianchi IX model in terms of Ashtekar's new
variables and how the solution to the supersymmetry equations (again
$\Psi = 1 $) is the state that Moncrief and Ryan showed to be
equivalent to that of Graham. This construction may allow to find
other states for different factor orderings and it also works for
other Bianchi models.  We will not focus our attention on the
supergravity methods.  However, one expects that by solving the
supergravity constraints in the Ashtekar's variables for the model in
question, only the state $\Psi = 1$ may be permitted.

The plan of the paper is as follows, in section II we
summarize the new variables for cosmology, the results of Ryan
and Moncrief and of Graham are presented in sections III and IV
respectively. In section V we draw the connection between them and
we end with a discussion of the possible implications of
these results.

\section{ NEW VARIABLES FOR BIANCHI MODELS: A SUMMARY}
In this section we briefly summarize the results presented in
\cite{Ko,AsPu}. The Ashtekar
new variables are a canonically conjugate pair consisting of a triad
$\tilde E_i^a$ (we denote densities with a tilde) and a complex $SO(3)$
connection $ A^i_a$.  The constraint equations become,
$$ {\cal G}^i = D_a \tilde E^{ai} \eqno(1) $$
$$ {\cal C}_b = \tilde E^{ai} F^i_{ab}\eqno (2) $$
$$ {\cal H}= \epsilon_{ijk} \tilde E^{ai} \tilde E^{bj}
F^k_{ab},\eqno (3) $$
where $F^i_{ab}$ is the curvature of $A^i_a$ and $D_a$ the covariant
derivative formed with $A^i_a$. An important point in this formulation
is that the variables are a priori complex. To retrieve real General
Relativity, one has to impose ``reality conditions''. One way of
imposing them is to require that the metric and its Poisson bracket
with the Hamiltonian be real.  This ensures that the resulting
formulation is equivalent to usual real General Relativity.

If one wants to restrict to Bianchi cosmologies one can separate the
time dependence of the variables and the fixed spatial dependence.
First introduce a fiducial basis of vectors $X^a_i$ and one forms
$\chi^i_a$ that implement the appropiate symmetry for the Bianchi
model in question $([X_{i},X_j]^a = C^k{}_{ij} X^a_{~k}$ and $ 2D_{[a}
\chi_{b]}^{~i} = - C^{i}{}_{jk} \chi_a^{~j} \chi_b^{~k})$. The indices
$a,b,...$ are spatial indices and the $i,j,...$ label the vectors and
forms in the basis and are raised and lowered with the Kronecker
delta. $ C^i_{jk} $ are the structure constants of Bianchi model in
question, $C^i{}_{jk} = 0 $ for Bianchi I, $C^i{}_{jk} =
\epsilon^i{}_{jk}$ for Bianchi IX.

In terms of these fixed bases we can expand the new variables as,
$$ \tilde E^a_j = E^i_j X^a_i \eqno (4) $$
$$ A^j_a = A^j_i \chi^i_a . \eqno (5) $$

In doing this, we concentrate all the spatial dependence in the
fiducial basis. The quantities $ E^i_j $ and $A^j_i$ are constants in
each three surface, that is, they only depend on ``time''.

Inserting these substitutions into the constraint equations \cite{foot1}
one gets,
$$ {\cal G}^i = C^k{}_{jk} E^{ij} + \epsilon^{ijk} A_{mj} E^m_k
\eqno (6) $$
$$ C_k = - E^j_i A^i_m C^m{}_{jk} + \epsilon^{imn} E^i_j A_{jm}
A_{kn} \eqno (7) $$
$$ {\cal H} = \epsilon^{ijk} C^p{}_{mn} E^m_i E^n_j A_{pk} +
E^m_i E^n_j (A^i_m A^j_n - A^i_n A^j_m ) . \eqno (8) $$

The reality conditions for Bianchi models read,
$$ q^{ij} = (q^{ij})^* \eqno (9) $$
$$ \dot q^{pq} = - iC^p_{mn} E^m_i E^n_j E^q_k \epsilon^{ijk}
+ 2 i E^i_j A^j_i q^{pq} + 2i E^p_i A^i_j q^{jq} +
m \leftrightarrow n =(\dot{q}^{ij})^*\eqno (10) $$
and can also be expressed in a nonpolynomial fashion by demanding
that $- A^{*i}_{~a} = A^i_a - 2 \Gamma^i_{~a}$, where
$ \Gamma^i_{~a}$ is the spin connection compatible with the triad.

If one further restricts to the diagonal models, the matrices
can be assumed to be diagonal, and the only remaining constraint
is the Hamiltonian\cite{foot2}. Introducing the following notation
for the variables,
$$ E^i_j = diag (E^1, E^2, E^3) \eqno (11) $$
$$ A^i_j = diag (A_1, A_2, A_3), \eqno (12) $$
the Hamiltonian constraint takes the form,
$$ {\rm Bianchi~ I:~ } {\cal H} = A_1 A_2 E^1 E^2 +
 A_1 A_3 E^1 E^3  + A_2 A_3 E^2 E^3 \eqno (13) $$
$$ {\rm Bianchi~ II:~ } {\cal H} = A_1 A_2 E^1 E^2 +
A_1 A_3 E^1 E^3 + (A_2 A_3 - A_1) E^2 E^3 \eqno (14) $$
$${\rm Bianchi~ VIII:~ } {\cal H} = (A_1 A_2 - A_3) E^1 E^2
+ (A_1 A_3 - A_2) E^1 E^3 + (A_2 A_3 + A_1 ) E^2 E^3
\eqno (15) $$
$$ {\rm Bianchi~ IX:~ } {\cal H} =  (A_1 A_2 - A_3 ) E^1 E^2
+ (A_1 A_3 - A_2) E^1 E^3 + (A_2 A_3 - A_1) E^2 E^3 . \eqno (16) $$

At this point one could consider the quantization of these
models. Start by choosing a realization, for instance
$\Psi [A]$, with
$$ {\hat E}_i \Psi [A] = { \partial \over \partial A_i}
\Psi [A] \eqno (17) $$
$$ {\hat A}_i \Psi [A] = A_i \Psi [A] \eqno (18) $$
and a factor ordering for the Hamiltonian constraint. For example
for the Bianchi IX model,
$$ {\hat {\cal H}} \Psi [A] = (A_1 A_2 - A_3)
{\partial^2 \over \partial A_1 \partial A_2} \Psi +
(A_2 A_3 - A_1) {\partial^2 \over \partial A_2 \partial A_3} \Psi
+ (A_1 A_3 - A_2) {\partial^2 \over \partial A_1 \partial A_3}.
\Psi \eqno (19) $$
In spite of the relatively simple appearance of this equation,
there are few ideas about how to construct the physical space
of states for the theory. Kodama explored some particular
solutions of this equation in reference\cite{Ko}. As can be seen, in
this factor ordering $\Psi[A]={\rm constant}$ is a solution.

The interesting point for our purpose however, is that the
constraint equations can be written in a unified fashion \cite{AsPu}.
Introducing the variables,
$$ Q^i_{~j} = E^i_k A^k_j \eqno (20) $$
the Hamiltonian constraint for all Bianchi class A models can
be written as,
$$ {\cal H} = Q^{*i}_{~~k} Q^k{}_i - Q^{*i}_{~~i} Q^j{}_j .
\eqno (21) $$

As can be readily seen, this version of the Hamiltonian constraint
does not have any explicit reference to the Bianchi model in
question! The dependence on the model appears in the
diffeomorphism constraint and in the symplectic structure for
the $Q$ variables.

The diffeomorphism constraint for diagonal models is identically
satisfied. This is important in connection with the unified
rewriting of the Hamiltonian constraint for all Bianchi models,
equation (21). Since the diffeomorphism constraint is not
present for diagonal models, all the dynamics of all class A
 diagonal models is summarized in equation (21), which
 particularizes to,
$$ {\cal H} = \overline Q_1  Q_2 + \overline Q_1 Q_3 +
\overline Q_2 Q_1 + \overline Q_2 Q_3 + \overline Q_3 Q_1 +
\overline Q_3 Q_2 = G_A^{~ij} \overline Q_i Q_j, \eqno (22) $$
where $i,j = 1...3$ and
$$ G_A^{~ij} = { 1 \over 2} \pmatrix { 0 & 1 & 1 \cr
 1 & 0 & 1 \cr 1 & 1 & 0 \cr } . \eqno (23) $$

The difference between different Bianchi models appears in
the sympletic structure and the reality conditions of the theory.

\section{THE RESULTS OF MONCRIEF AND RYAN}
Moncrief and Ryan \cite{MoRy} recently explored amplitude-real-phase
exact solutions to the quantum Hamiltonian constraint in terms
of the more traditional Misner-type variables. We briefly summarize
their results here.

The Misner type variables are obtained by parametrizing ~the ~three
metric and for the dia\-go\-nal ca\-ses
ta\-king the ma\-trix $\beta_{ij}$ as
$\beta_{ij} = diag (\beta_+ + \sqrt {3} \beta_-, \beta_+ - \sqrt {3}
\beta_-, -2\beta_+)$. In terms of the variables $(\alpha, \beta_+,
\beta_-)$ and their conjugate momenta $ (p_\alpha, p_+, p_-)$ the
Hamiltonian constraint becomes,
$$ {\cal H} = \exp(-3\alpha) (-p^2_\alpha + p^2_+ + p^2_-  +
\exp(4\alpha) V(\beta_\pm)) , \eqno (24) $$
where $V(\beta_\pm)$ is a function that depends on the particular
type of Bianchi model consdidered. For the Bianchi IX case it
is given by,
$$ V(\beta_\pm) = {1 \over 3} \exp(-8 \beta_+) - {4 \over 3}
\exp(-2\beta_+) \cosh(2 \sqrt {3} \beta_-) + {2 \over 3}
\exp(4 \beta_+) (\cosh(4 \sqrt {2} \beta_-) -1). \eqno (25) $$

Equation (24) can be rewritten in an enlightening form,
$$ {\cal H} = G^{ij} p_i p_j + U(q) , \eqno (26) $$
where the variables $q^i, \quad i = 1,2,3$ refer to $\alpha, \beta_\pm$ and
$p_i$ their corresponding momenta. $G^{ij} $ is just the flat
Minkowski metric in 2+1 dimensions. One can therefore interpret, by
considering $q^0 = \alpha$ to be a time coordinate,
the dynamics of the Bianchi
cosmologies as that of a massless particle moving in a (time
dependent) potential  in 2 + 1 dimensions. This permits a
good qualitative understanding of the dynamics of the Bianchi
cosmologies.

If one quantizes the model taking a representation
$\Psi (\alpha, \beta_\pm)$, there is a factor ordering ambiguity
in the first term of the Hamiltonian constraint. Hartle and Hawking
\cite{HaHa} suggested the following ``semigeneral'' factor ordering,
$-\exp(-3 \alpha) \partial^2_\alpha +
B \exp(-3\alpha)\partial_\alpha $ with $B$
an arbitrary constant. The resulting quantum constraint
 (WDW equation) is,
$$\bigg ( - {\partial^2 \over \partial \alpha^2} +
{\partial^2 \over  \partial \beta^2_+ } +
{\partial^2 \over \partial \beta_-} +
B {\partial \over \partial \alpha} -
\exp(4\alpha) V(\beta_\pm) \bigg ) \Psi = 0 \eqno (27). $$
Moncrief an Ryan set out to find solutions of the form
$$ \Psi [\alpha, \beta_\pm ] = W(\alpha, \beta_\pm)
\exp(-\Phi(\alpha, \beta_\pm)  ). \eqno (28) $$

If one inserts this ansatz into equation (27) one gets a
complicated partial differential equation for $\Phi$ and
$W$. Moncrief and Ryan make a further choice requiring
that $\Phi$ satisfy,
$$  \bigg ( {\partial \Phi \over \partial \alpha} \bigg )^2 -
\bigg ( {\partial \Phi \over \partial \beta_+ } \bigg )^2 -
\bigg ({\partial \Phi \over \partial \beta_- } \bigg )^2 +
\exp (4 \alpha) V(\beta_\pm) = 0 . \eqno (29) $$

Finding a solution to equation (29) one is left with solving a
PDE for $W$. A solution to (29) can actually be found of the
form,
$$ \Phi = {1 \over 6} \exp(2\alpha)(\exp(-4\beta_+) +
2 \exp(2\beta_+) \cosh(2\sqrt {3} \beta_-)) \eqno (30) $$
and with this ansatz for $\Phi$, choosing $B= -6, W = $ const.
is a solution. What Moncrief and Ryan really demanded is
$\Phi$ to be the solution of the ``Euclidean'' Hamilton-Jacobi equation (29)
for this particular model.

What does this have to do with Ashtekar's variables? Kodama proved
that if one takes a wavefuntion in terms of Ashtekar's variables
$\Psi_A$, one can reconstruct the wavefunction in terms of the
traditonal variables by choosing $\Psi_{traditional} = \exp(\pm i
\Phi_A) \Psi_A$, with $\Phi_A$ given by $\Phi_A = 2 i \int \tilde E^a_i
\Gamma^i_{~a} d^3 x$. For the Bianchi IX case, $\Phi_A$ is equal to
the $\mp i\Phi$ found by Moncrief and Ryan and therefore the solution
$\Psi_A$ = const in terms of Ashtekar variables
becomes, when transformed to the traditional
variables, the wavefuction found by Moncrief and Ryan by direct
substitution.

\section{THE WAVEFUNCTION OF GRAHAM}

A totally (apparently, as we will see) independent result
was found by Graham \cite{Gr}. He considered a supersymmetrization of
the Bianchi system using the fact that he was able to solve
the ``Euclidean'' Hamilton-Jacobi equation (29) which he
writes in the form,
$$ U(q) = G^{ij} {\partial \phi \over \partial q^i} {\partial
\phi \over \partial q^j} , \eqno (31) $$
where $\phi $ is the $\Phi$ of equation (29). He finds the solution
(30), the same as that given by Moncrief and Ryan.  That this
decomposition can be accomplished at all can be readily seen if one
writes the potential in terms of the metric variables without using
the Misner parametrization. In terms of those variables the
potential term is just a second order polynomial.

With this observation, one can
introduce a set of fermionic variables $\psi^i , \overline
\psi^j$ satisfying the spinor algebra,
$$\psi^i \psi^j + \psi^j \psi^i = 0 \eqno (32) $$
$$\overline \psi^i \overline \psi^j + \overline \psi^j
\overline \psi^i = 0 \eqno (33) $$
$$ \overline \psi^i \psi^j + \psi^i \overline \psi^j =
G^{ij} \eqno (34) $$
and define the following supercharges,
$$ {\cal S} = \psi^i \bigg (p_i + i {\partial \phi \over
\partial q^i} \bigg ) \eqno (35) $$
$$ \overline {\cal S} = \overline \psi^i \bigg ( p_i - i
{\partial \phi \over \partial q_i} \bigg ) , \eqno (36) $$
which satisfy ${\cal S}^2 = \overline {\cal S}^2 = 0 $,
such that the Hamiltonian can be written in the following
form
$$ {\cal  H} = {1 \over 2} ({\cal S} \overline {\cal S} +
 \overline {\cal S} {\cal S} ) . \eqno (37) $$

So we see the ${\cal S}$'s work as ``square roots'' of the
Hamiltonian. A quantum representation can also be introduced
considering wavefuctions of the canonical coordinates and of three
Grassmann variables $\eta^i$ in terms of which one can represent the
$\psi$ algebra by, $\psi^i = \eta^i, \overline \psi^i = G^{ij} {\delta
\over \delta \eta^j} $.  However if one writes the quantum Hamiltonian
as ${\hat {\cal H}} = {1 \over 2} ({\hat {\cal S}} {\hat {\overline
{\cal S}}} + {\hat {\overline {\cal S}}} {\hat {\cal S}} ) $ there is
a factor ordering discrepancy from simply quantizing the traditional
Hamiltonian by a term $\hbar { \partial^2 \phi \over \partial q^i
\partial q^j} [ \overline \psi^i, \psi^j ]$.

Any solution of ${\hat {\cal H}} \Psi = 0 $ can be written as $$ \Psi
= A_+ + B_i \eta^i + C^k \epsilon_{kij} \eta^i \eta^j + A_-
\epsilon_{ijk} \eta^i \eta^j \eta^k , \eqno (38) $$ where the eight~
fuctions $ A_\pm, B_i, C^k$ depend ~on the canonical~ configuration
variables $q^i = q^i (\alpha, \beta_\pm) $.  In the non-supersymmetric
(bosonic) limit $A_+$ should be a solution of the traditional
WDW equation. Therefore if one is able to find a $\Psi$
solution to this model implicitly one finds a solution to the quantum
constraints of the usual Bianchi IX cosmology.

Wavefunctions should be supersymmetric, that is they should
satisty ${\hat S} \Psi = 0 ,  {\hat {\overline S}} \Psi = 0$.
One can solve these equations in general for the bosonic part
of the wavefunction. One gets,
$$ A_+ = a_+ \exp(-\phi/\hbar) \eqno (39) $$
where $a_+$ is a constant.

This in particular implies that this quantity should be a
solution of the usual WDW equation for the Bianchi
IX cosmology. This is readily seen. Recalling the form of $\phi$,
one sees that this wavefunction is modulo irrelevant constants
the same as the one found by Ryan and Moncrief.

The reader may be surprised that in this formulation this appears as
the general (pure bosonic) solution: of course, when one sees that
equation (31), is the same as equation (29), different boundary
conditions will give different solutions for $\phi$ (for the Taub
model, that is, a diagonal Bianchi IX model with $\beta_- = 0$, there
exist two distinct solutions for $\phi \cite{GuRoRy}$). The procedure of
Graham will then lead to a family of solutions, one for each
$\phi$. In fact they have been found, see \cite{Ko}. What one
encounters here is a common phenomenon in supersymmetry, that is, the
kernel of an operator is usually smaller than that of its square.

In particular when, supergravity is used the constraint
equations for the Bianchi models are so restrictive that
only the fuctions~ $A_+$ and ~$A_-$ remain as parts of the
wave equation (38). ~They ~are the~ quantum states ~one
would obtain
by~ path integral ~methods \cite{DeHaOb,De}.

Moreover, not all solutions to the Hamilton-Jacobi equations
will be consistent with the desired boundary conditions and
also the supergravity constraints \cite{De2}. However, we will not
focus our attention on these issues here.

\section{SUPERSYMMETRIZING THE ASHTEKAR FORMULATION}

In spite of the fact that both the construction of Moncrief and Ryan
and that of Graham make use of the same Hamilton-Jacobi function
$\phi$, it is somewhat remarkable that such radically different
techiques yield the same wave function. Here we will seek to put them
on a common ground.  Let us start by recalling that in terms of the
Ashtekar variables the Hamiltonian constraint can be written as, $$
{\cal H} = \overline Q_1 Q_2 + \overline Q_1 Q_3 + \overline Q_2 Q_1 +
\overline Q_2 Q_3 + \overline Q_3 Q_1 + \overline Q_3 Q_2 . \eqno (40)
$$ This is a ``universal constraint''which is insensitive to the
choice of a specific diagonal model. The dependence on the specific
structure constants of the Bianchi model under consideration will
appear by imposing the reality conditions.

This Hamiltonian inmediately suggests to introduce the
supercharges,
$$ {\cal S} = \psi^i Q_i \eqno (41) $$
$$\overline {\cal S} = \overline \psi^i Q^*_i . \eqno (42) $$

The algebra of the $\psi$'s is the same as before except that now,
$$ \overline \psi^i \psi^j + \psi^i \overline \psi^j = G_A^{~~ij}
\eqno (43) $$
with $G_A$ given by equation (23).

In terms of these supercharges we can again write the
Hamiltonian constraint as $ {\hat {\cal H}} = {1 \over 2}
({\hat {\cal S}} {\hat {\overline {\cal S}}} +
{\hat {\overline {\cal S}}} {\hat {\cal S}})$. ~Also ~we
can introduce ~a quantum~ representation.~ Since the ~$Q$ variables
 are noncanonical, we will choose the more usual ``connection''
 representation in terms of fuctions of $A_i$ and
 $\eta^i$ and we could seek for the bosonic part of the
 solution to the supersymmetry constraints. The equations become,
$$ {\hat {\cal S}} \Psi [A,\eta] = {\hat Q}_i \eta^i
\Psi [A, \eta] \eqno (44)$$
$$ {\hat {\overline {\cal S}}} \Psi [A, \eta] = {\hat Q }^*_i
{\partial \over \partial \eta^i} \Psi [A, \eta] \eqno (45) $$
and using the reality conditions in terms of the Ashtekar
connection, $- A^{*i}_{~~a} = A^i_a - 2 \Gamma^i_a$ we get,
$$\eta^i A_i {\partial \over \partial A_i} \Psi [A, \eta] =
0 \eqno (46) $$
$$ (A_k - 2 {\hat \Gamma}_k ) {\partial \over \partial A_k }
G^{ki}_A {\partial \over \partial \eta_i} \Psi [A, \eta] = 0 .
\eqno (47) $$

The only solution that this system of equations admits is
$\Psi[A]={\rm constant}$.  This may seem a trivial solution but as we
saw in previous subsections, it is exactly the kind of solution that
Ryan and Moncrief showed to have a very nontrivial form in terms of
the traditional variables. It then turned that it was equivalent to
the solution that Graham found for the bosonic sector,
supersymmetrizing the traditional formulation and the only component
permitted by supergravity \cite{De2}.

Notice that the supersymmetrization we have performed works
for arbitrary Bianchi models. This seems quite natural,
and in fact the Graham construction has already been
generalized to the Bianchi II model \cite{ObSoBe}, where the whole
wavefunction was obtained.

\section{CONCLUSIONS}
We have therefore shown that the Ashtekar variables~ provide a
natural framework for
seeking supersymmetric quantum states. We find that the only possible
bosonic state is a constant, which translated in terms of the
traditional variables corresponds to the result that Graham found, as
Ryan and Moncrief proved and is also connected to the state obtained
by means of supergravity \cite{DeHaOb,De}. This state could be
understood as the wormhole quantum state \cite{De,HaPaVi,De2}.

What does this tell us about the general theory?  One in general
cannot formulate the Hamiltonian constraint in the Ashtekar
formulation in terms of the $Q$ variables.  However taking a
particular lapse gauge a similar formulation is possible and has found
application in the asymptotically flat context \cite{Asbi}. There is
however, the issue of the other constraints (diffeomorphism and gauss
law) to take care of. It would be interesting to pursue this line of
reasoning further to see if it in some sense it simplifies finding
solutions to the generic WDW equation. It is however, at
first sight a bit disappointing to notice that even in the Bianchi
context the only solution the technique yields is a constant. If one is
to take seriously that somehow supersymmetry selects a preferred
ground state \cite{DeHaOb} this could tell us something about the
ground state of the general theory \cite{De2}.

What does this formulation have to do with supergravity?  In principle
it is a different construction, i.e.  supersymmetrizing a Bianchi
model as a mechanical system is not the same as particularizing
supergravity for a minisuperspace. It would however be interesting to
compare both cases in terms of the new variables (supergravity has
already been formulated in terms of them by Jacobson \cite{Ja}) and
study similarities and differences. On the other hand, again one could
study the implications for the general theory.  For non-zero
cosmological constant, a semi-classical WKB wave function \cite{SaSh}
has been obtained for the full supergravity (N=1) in Ashtekar's
formalism it has the form of the exponential of the Chern-Simons
functional and has been particularized for the Robertson-Walker
universe.  Moreover the general, supergravity theory N=1 seems to
select out the most ~symmetrical ~states~ \cite{DeHaOb}, ~the ~whole
the standard variables, the two expected bosonic states $\exp (\pm I)
$ appear and all physical states are given by finite expressions.  It
is mostly interesting to look if these results can be obtained in
terms of the Ashtekar variables and try to understand them in
connection with the ``ground state'' of the theory.

\section*{Acknowledgements}
This work was supported in part by CONACyT Grants
F246-E9207 and 1683-E9209, funds provided by the University of Utah
and NSF grant PHY92-07225. J.P. wishes to thank the organizers of the
Vth Mexican School of Particles and Fields, The Institute for
Theoretical Physics at UCSB and grant NSF-PHY89-04035 for hospitality
and support during stages of this work.


\begin{references}
\bibitem{foot1} We will only be concerned with Class A
models (I,II,VIII,IX). Class B models don't admit
standard Hamiltonian treatments and must be considered
separately.
\end{references}
\end{document}